\documentclass[letterpaper,pra,twocolumn,showpacs,superscriptaddress]{revtex4}

\usepackage{graphicx}
\usepackage{amsmath}
\usepackage{amssymb}

\begin{document}

\title{Experimental Demonstration of Squeezed State Quantum Averaging}


\author{Mikael Lassen,$^{1,*}$ Lars Skovgaard Madsen$^1$, Metin Sabuncu$^{2}$, Radim Filip$^{3}$, and Ulrik L. Andersen$^{1}$}

\address{Department of Physics, Technical University of Denmark, 2800 Kongens Lyngby, Denmark}
\address{Max-Planck-Institute for the Science of Light, G\"{u}nther-Scharowsky-Str. 1, 91058 Erlangen, Germany}
\address{Department of Optics, Palack\' y University, 17. listopadu 50,  772~07 Olomouc, Czech Republic}
\email{mlassen@fysik.dtu.dk}

\date{\today}

\begin{abstract}
We propose and experimentally demonstrate a universal quantum averaging process implementing the harmonic mean of quadrature variances. The harmonic mean protocol can be used to efficiently stabilize a set of fragile squeezed light sources with statistically fluctuating noise levels. The averaged variances are prepared probabilistically by means of linear optical interference and measurement induced conditioning. We verify that the implemented harmonic mean outperforms the standard arithmetic mean strategy. The effect of quantum averaging is experimentally tested both for uncorrelated and partially correlated noise sources with sub-Poissonian shot noise or super-Poissonian shot noise characteristics.
\end{abstract}


\maketitle

The mean of a set of statistically varying, real and non-negative numbers, $x=\{x_1,x_2,...x_n\}$, is in general~\cite{Krenz2000.book}:
\begin{equation}
M(x)=\left(\frac{1}{n}\sum_{i=1}^n{x_i^r}\right)^{1/r},
\end{equation}
where $r$ is an integer. The most commonly used means are the arithmetic mean and the harmonic mean corresponding to $r=1$ and $r=-1$, respectively \cite{Bullen1987.book}. These two kinds of means occur in several physical problems, and the actual mean being used to describe the physical system depends on the physical setup. E.g. the total resistance, $R$, of an electrical circuit consisting of $n$ serially or parallelly connected resistors (with resistances $R_i$) is known to follow the arithmetic, $R=\sum_{i=1}^n R_i$, and harmonic, $\frac{1}{R}=\sum_{i=1}^n\frac{1}{R_i}$, means, respectively (apart from a multiplicative constant). Similar mean laws can also be deduced for the total stiffness of a system comprising springs connected in series or in parallel. Likewise, one finds examples of the arithmetic mean and the harmonic mean in geometrical optics as well as in astronomy~\cite{Fundamentals2007.book,Astr.book}.

All these examples of the arithmetic and harmonic means are based on {\it classical} systems. In this Letter, we explore an example of the arithmetic and the harmonic mean in a {\it quantum} optical system. More specifically, we propose and experimentally demonstrate the arithmetic and harmonic means of the quadratures variances of different quantum states using an optical system that is based on simple linear optics and homodyne detection. We average an ample supply of quantum states that exhibit either sub-Poissonian shot noise or super-Poissonian shot noise behavior, and we investigate the averaging procedure for completely independent as well as partially noise correlated quantum states. Such averaging protocols allows us to stabilize the degree of squeezing (variances) of $n$ independent, fragile and possibly unstable squeezed state resources. Therefore, besides being of fundamental interest, such a protocol will find applications in quantum information and quantum metrology where stable resources of squeezed light are required~\cite{Andersen_cvQuInPro,Giovannetti2006}. We find that in terms of stabilizing the squeezed state variance from a fluctuating set of different noisy squeezed light resources, the harmonic mean outperforms the standard arithmetic mean.

\begin{figure}[th]
\begin{center}
\includegraphics[width=0.35\textwidth]{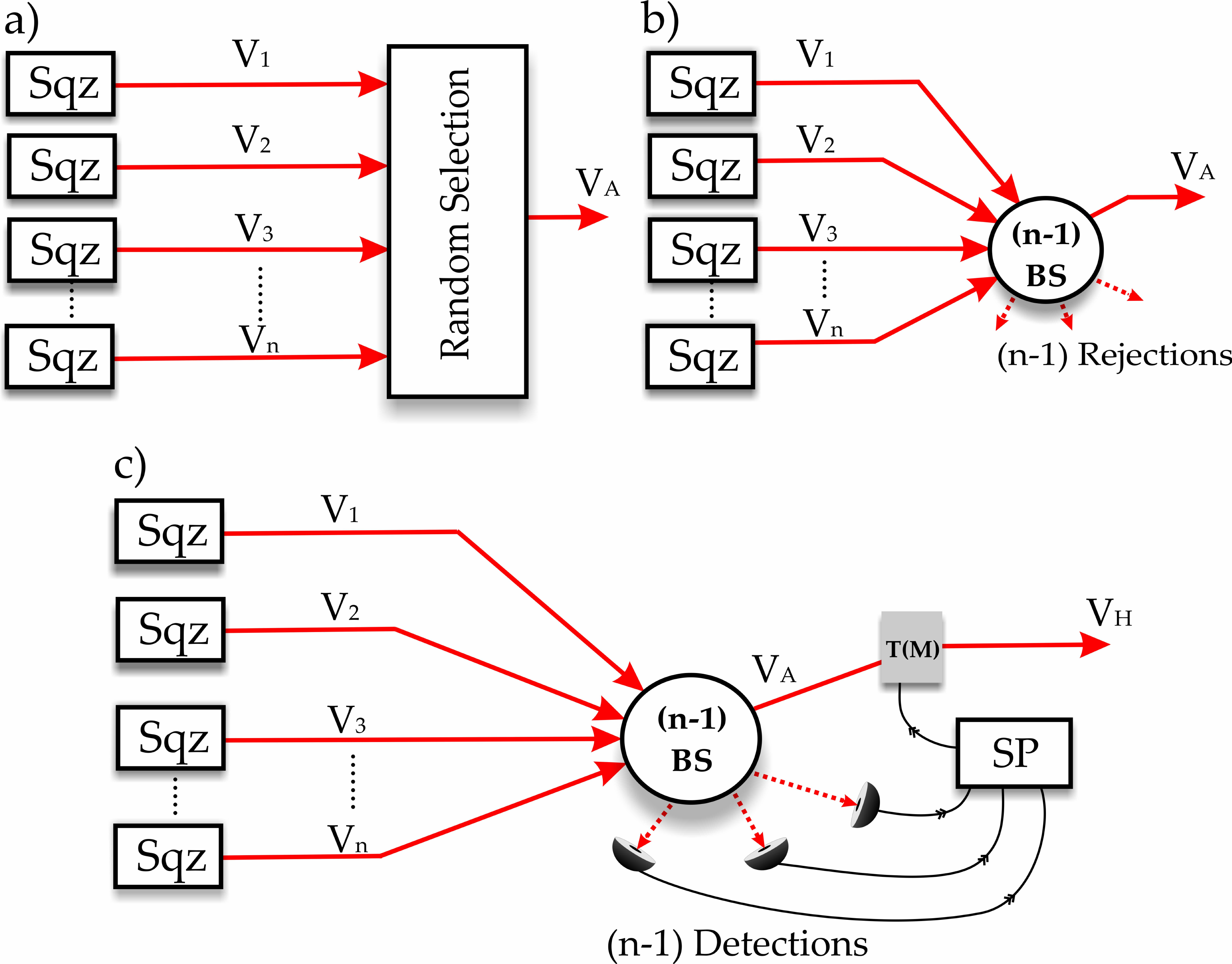}
\caption{Schematic illustrations of two arithmetic mean protocols and the harmonic mean protocol (see text for details). SP: Signal processing; BS: Beam splitters; SQZ: Squeezing source; T(M): Trigger (modulator). Experimentally we explore setup b) and c).}
\label{fig1}
\end{center}
\end{figure}

Consider $n$ independent quantum resources described by the quadrature variances $V_i=\langle x_i^2\rangle-\langle x_i\rangle^2$ where $x_i,\; (i=1,2...n$) are the amplitude quadratures. The goal is to construct the arithmetic and harmonic means of the variances without using any additional squeezed state resources. The arithmetic mean can be formed in two different ways as illustrated in Fig.~\ref{fig1}: a) By randomly selecting one of the $n$ states from the resources or b) by interfering the $n$ states on $(n-1)$-beam splitter~\cite{reck1994} and subsequently rejecting all outputs except one. Any one of these approaches yields the arithmetic mean for the total variance:
\begin{equation}
V_A=\frac{1}{n}\sum_{i=1}^n V_i.
\end{equation}
Note that such a strategy is similar to the demonstration of universal quantum purification of continuous variable quantum information based on an ensemble of identical states~\cite{Andersen2005}.

The harmonic mean of the variances can be formed by using the setup shown in Fig.~\ref{fig1}c. Here the $n$ states interfere on ($n-1$)-beam splitter, the amplitude quadratures of the ($n-1$) outputs are measured and the results are used to drive the state. The state can be driven in two different ways: If the amplitude quadrature variances of the input states are known, the protocol is deterministic and the state is linearly displaced with an amount determined by the a priori variance information as well as the measurement outcomes. However, in a more relevant scenario where the variances are unknown, the state is probabilistically heralded based on the measurement outcomes: If the outcomes are arbitrarily close to zero, the state is kept, otherwise it is discarded. Importantly, the latter protocol is {\it universal} as it is independent on the input state. Either methods yield the harmonic mean for the variances:
\begin{equation}
\frac{1}{V_H}=\frac{1}{n}\sum_{i=1}^n \frac{1}{V_i}.
\end{equation}
We note that a quantum optical version of the "resistor"-type harmonic mean for the amplitude quadratures, $1/V_H=\sum_{i=1}^n1/V_i$, can be also implemented exactly solely using Gaussian operations, by replacing the array of beam splitters in Fig.\ref{fig1}c with quantum non-demolition interactions. This, however, requires the use of additional squeezing resources and is therefore not considered further in this paper.

\begin{figure}[th]
\begin{center}
\includegraphics[width=0.4\textwidth]{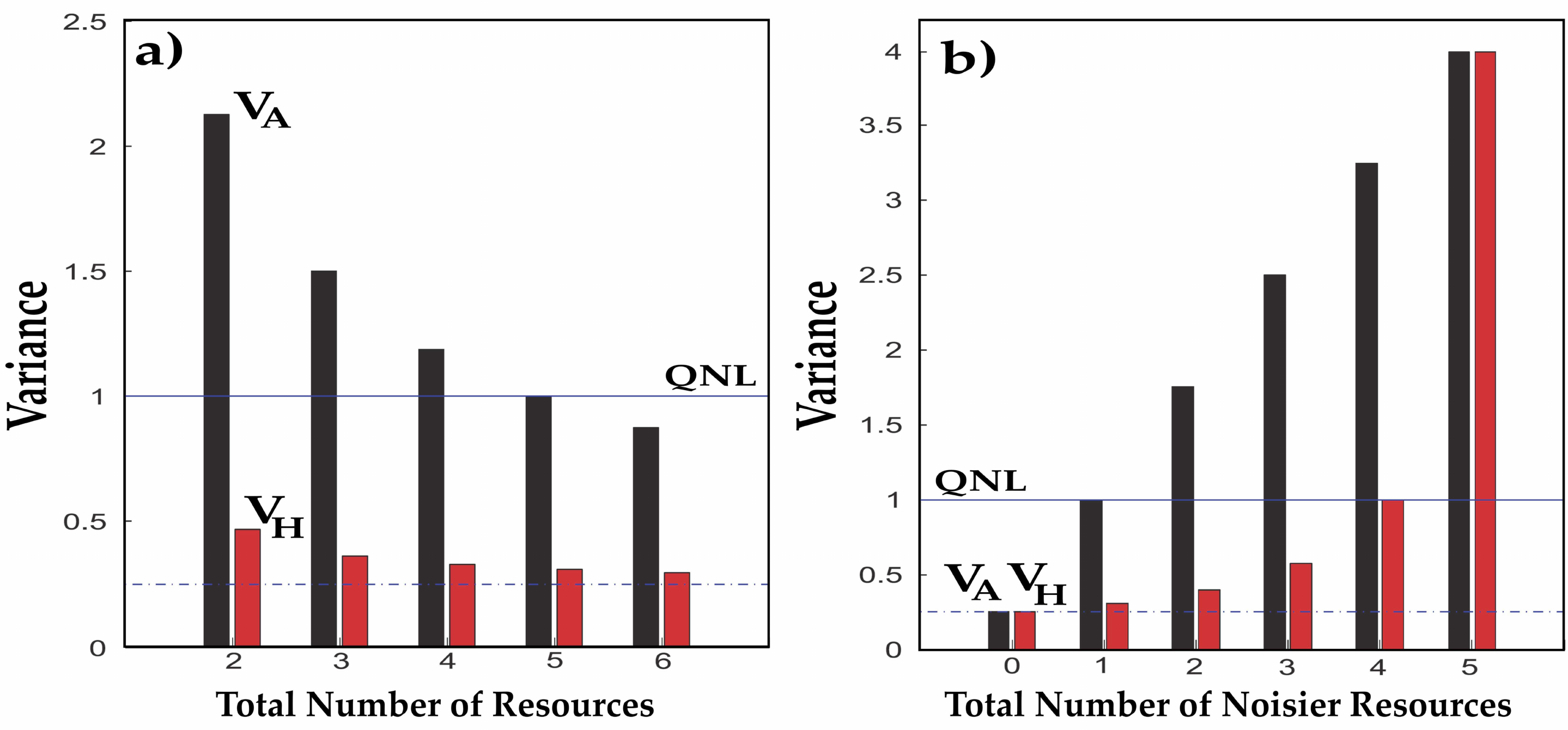}
\caption{Comparison between the two different means. Variances of the arithmetic and harmonic means a) for an increasing number of resources (one with variance $V=4$ and $V=0.25$ for the rest), and b) for a supply of five resources with an increasing number of noisy resources (with $V=4$) and the rest being quiet with $V=0.25$.}
\label{fig2}
\end{center}
\end{figure}

By using the two averaging operations, it is possible to stabilize the variances of $n$ independent unstable squeezed state resources. As an example, let us consider four noise resources with $V=0.25$ and a single broken source with $V=4$ (both variances are normalised to the variance of a vacuum state). The arithmetic mean transformation produces a single source with $V_A=1$ whereas the harmonic mean transformation produces a source with $V_H=0.31$. Interestingly, the arithmetic mean protocol stabilizes the sources to the shot noise level and the harmonic mean stabilizes it below the shot noise level. Moreover, if another source is broken (also with $V=4$), then one gets an arithmetic mean of $V_A=1.75$ and a harmonic mean of $V_H=0.40$, which means that the harmonic mean method is much less sensitive to the number of broken resources. The two means are compared in Fig.~\ref{fig2}a for different numbers of resources and a single noisy resource with $V=4$, and in Fig.~\ref{fig2}b for different numbers of noisy resources with a total of 5 resources.

We note that a stabilization procedure for an ample supply of continuous variable quantum informational states as well as qubit transformations were addressed theoretically in ref.~\cite{Marek2007} and refs.~\cite{Berthiaume1994,Peres1999}, respectively. Unlike these proposals, in our work we stabilize off-line resources without disturbing the actual quantum processor. Furthermore we note that in contrast to Gaussian squeezed state distillation (which is not possible using linear optics~\cite{Kraus.2003}), quantum averaging can be implemented with solely linear optical elements and feedforward. In addition, we stress that the squeezed state distillation of Non-Gaussian noise in refs. \cite{Heersink2006,Franzen2006} was state dependent whereas our protocol is universal.

\begin{figure}[t]
\begin{center}
\includegraphics[width=0.4\textwidth]{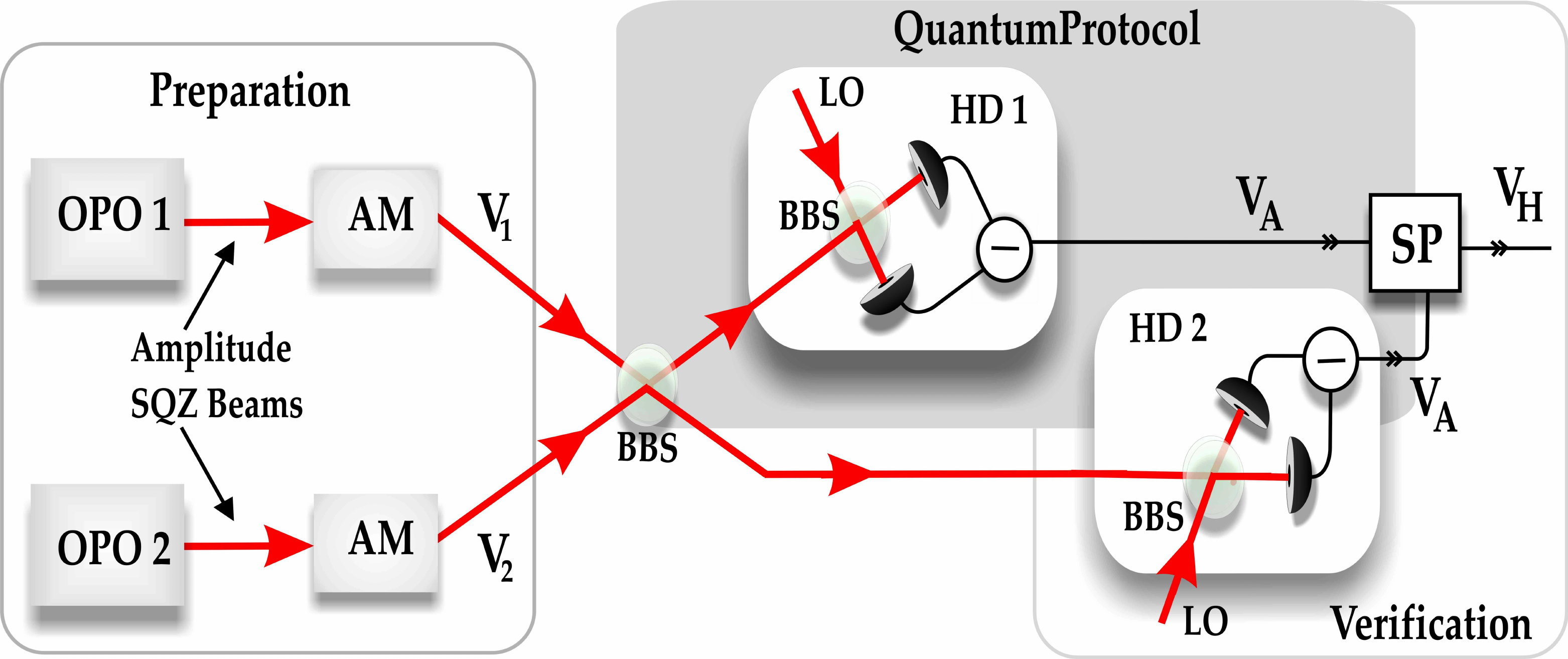}
\caption{Schematic of the laboratory setup for the implementation of quantum averaging. Squeezed states are prepared in optical parametric oscillators (OPOs) and additional noise can be added in amplitude modulators (AMs). The measurement induced harmonic mean operation and the verification are performed with homodyne detectors (HD1 and HD2). $V_1$ and $V_2$ are the input variances, $V_A$ is the arithmetic mean and $V_H$ is the harmonic mean. BBS: Balanced beams splitter; SP: Signal processing; LO: Local oscillator.}
\label{fig3}
\end{center}
\end{figure}

We experimentally implement the two averaging transformations for a supply of two quantum states ($n=2$) that exhibit either sub-Poissonian shot noise or super-Poissonian shot noise behavior. The schematics of our experiment is depicted in Fig.~\ref{fig3}. Basically, the setup comprises three parts; a quantum state preparation part, a quantum protocol part and a verification part. We prepare two Gaussian states using optical parametric oscillators (OPO) followed by amplitude modulators (AM). The OPOs are bow-tie shaped cavities with type I periodically poled KTP crystals~\cite{Huck2009,Lassen2009}. To produce a bright excitation of the squeezed beam and to enable a cavity phase lock, we inject two auxiliary beams into the cavity. Squeezed states are produced at 1064~nm by pumping the parametric process with a mode-matched beam at the wavelength of 532~nm. The relative phases between the pump and seed beams are locked to de-amplification in order to generate amplitude squeezed beams. To produce a whole range of different noise variances, the amplitude squeezed states are sent through amplitude modulators (AM) which are driven by electronic noise sources with variable modulation depths. Using such a combination of OPO and AM, the noise of the amplitude quadratures can be tuned from sub-Poissonian shot noise to super-Poissonian shot noise. To execute the protocol, the resulting beams interfere (with a visibility higher than 98\%) on a balanced beam-splitter (BBS). The relative phase is set such that the two beams add in quadratures. The two output beams of the beam-splitter are then measured by homodyne detectors (HD) with slowly varying local oscillator phases. The measurements are performed at the sideband frequency of 4~MHz with a bandwidth of 300~kHz, and the output signals were amplified and digitized at $5 \times 10^6$ samples per second. Data bins associated with the measurements of the amplitude quadratures contains approximately $6 \times 10^4$ data points.

The arithmetic mean is produced directly after the beam splitter by discarding the outcomes of HD1. To implement the harmonic mean protocol, we select the outcomes of HD2 based in the outcomes in HD1 with varying threshold values: If the data point of HD1 is lower than the threshold value, the corresponding data point of HD2 is kept, otherwise it is discarded. The state heralding process could in principle also be implemented electro-optically to generate a freely propagating averaged quantum state. However, to avoid such complications, our conditioning is based on digital data post-selection.

\begin{figure}[t]
\begin{center}
\includegraphics[width=0.45\textwidth]{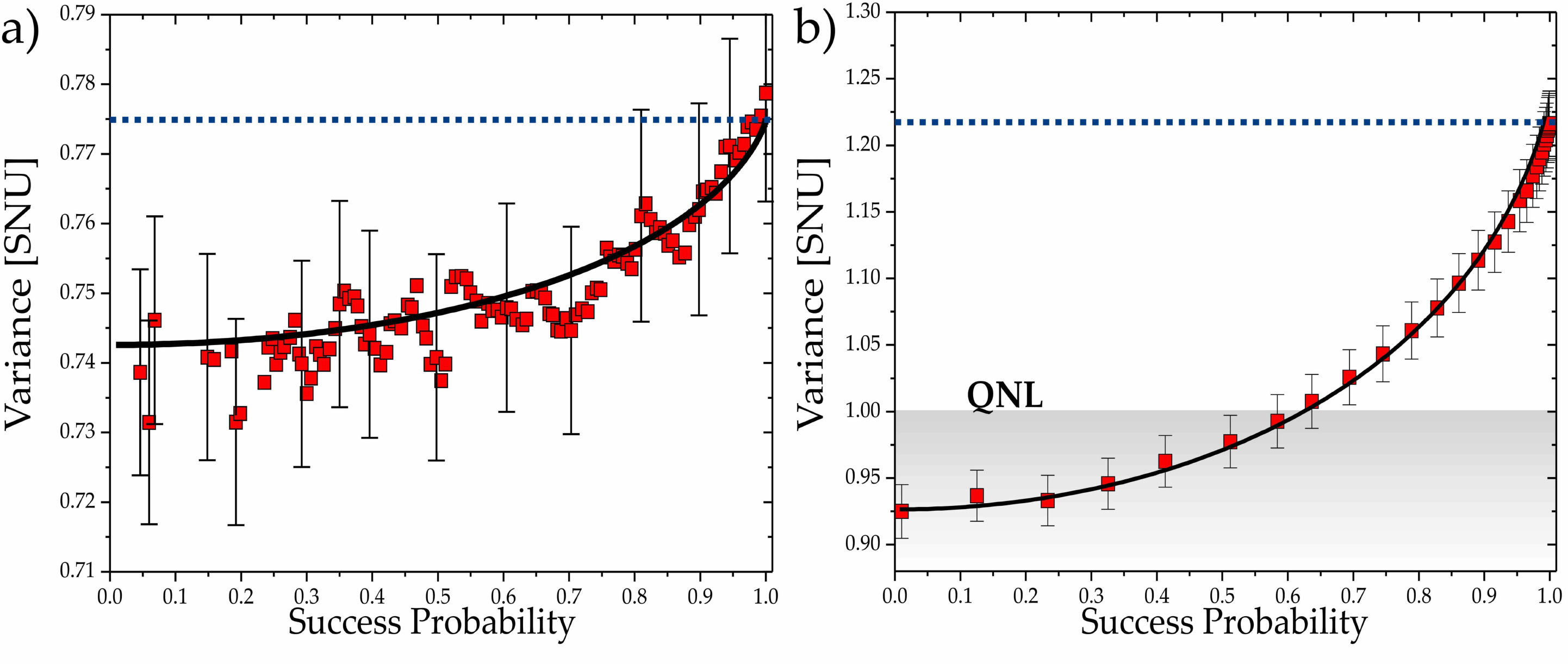}
\caption{The arithmetic mean and harmonic mean as a function of the success rate $P_S$ for for two different supplies of resources. In a) the two input states are squeezed with variances $V_1=0.64\pm0.01$ and $V_2=0.90\pm0.02$, and in b) only one of the resources are squeezed with variance $V_1=0.62\pm0.01$ while the other resource has a variance $V_2=1.83\pm0.02$ above the quantum noise level (QNL). Theoretical predictions are represented by solid (harmonic mean) and dashed (arithmetic mean) curves.}
\label{fig4_expData}
\end{center}
\end{figure}

We present the experimental results for the arithmetic and harmonic mean protocols in Fig.~\ref{fig4_expData}. Fig.~\ref{fig4_expData}a shows the results for two amplitude squeezed beams with variances $V_1=0.64\pm0.01$ and $V_2=0.90\pm0.02$. We see that for a success probability (ratio of the data kept after post selection to the initial data) of around $0.10$, the harmonic-mean method produces a state with $V_{H}=0.74\pm0.02$ whereas the arithmetic-mean method gives $V_{A}=0.77\pm0.02$. In this particular case the improvement of the amplitude noise using the harmonic mean method is very small and is basically within the measurement uncertainty. However, the superior performance of the harmonic mean with respect to the arithmetic mean is clearly manifested by considering the input variances of $V_1=0.62\pm0.01$ and $V_2=1.83\pm0.04$. As shown in Fig.~\ref{fig4_expData}b, in this case the arithmetic-mean method produces a state with $V_{A}=1.22\pm0.03$ whereas the harmonic mean method produces a squeezed state for success probabilities lower than $0.60$. With a success probability around $0.10$, the harmonic-mean method generates a state with $V_H\approx 0.90\pm0.03$. This shows that the universal harmonic mean strategy can stabilize very fragile and unstable quantum noise sources below the quantum noise limit (QNL) against a source that suddenly generates a large amount of classical noise. The solid curves in Fig.~\ref{fig4_expData} represents the theory taking into account various imperfections of the setup. The error bars of the variances depend on the measurement error, which mainly are associated with the stability of the system over time. During a complete measurement (lasting 3-4 hours) the standard deviation of the quantum noise level was 0.02 shot noise units. And the statistical errors which are due to the finite measurement time and the post selection process~\cite{frieden83.book}.

In the above analysis and experiment, the $n$ sources are completely independent and thus the amplitude quadratures are uncorrelated, $C=\langle X_1....X_n\rangle=0$. We now consider the situation of partially correlated sources. Although the theory can be easily conducted for an arbitrary number of sources, for simplicity we consider only the case of two sources ($n=2$) with a quadrature correlation described by the coefficient $C=\langle X_1X_2\rangle$. Using the same setups as above, we find that the arithmetic mean is modified to $V_{Ac}=\frac{V_1+V_2}{2}-C$, whereas the probabilistic harmonic mean is approaching $V_{Hc}=2\frac{V_1 V_2-C^2}{V_1+V_2+2C}$ for a very narrow post selection interval. Through simple inspection, we see that the effect of the correlations ($C>0$) is a reduction of the mean values. Moreover, we easily find that $V_H=V_{Hc}$ if the correlations are unbiased and otherwise $V_H>V_{Hc}$. We also stress that these protocols are, as above, universal in the sense that no a priori information about the input state is required to execute the transformations. On the other hand, if the noise is maximally correlated and unbiased (and this is a priori known), the noise can be removed perfectly and deterministically as shown in ref.~\cite{Bowen}.

\begin{figure}[t]
\begin{center}
\includegraphics[width=0.45\textwidth]{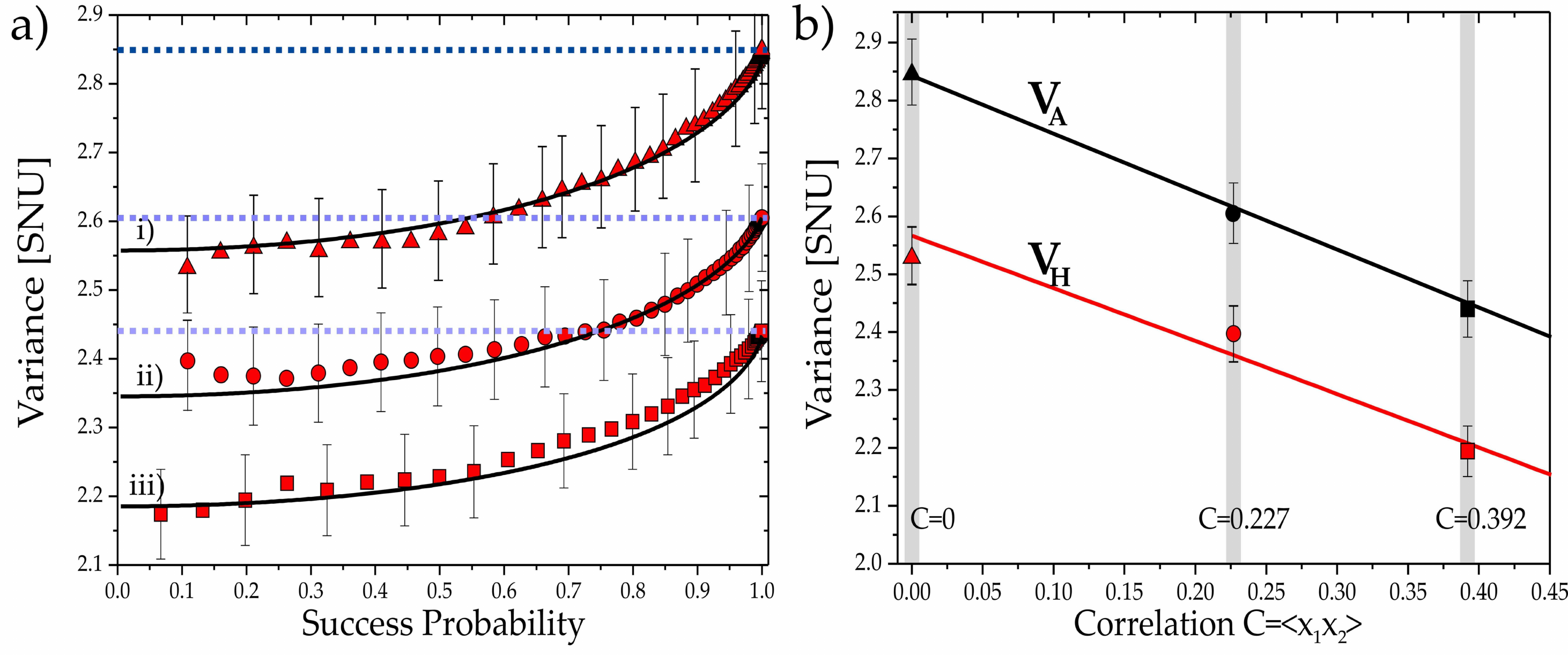}
\caption{The influence of correlations between the resource states. In a) we plot the variances as a function of the success probability for different degrees of correlation and the results are summarized in b) for a success probability of 10\%. The variances of the input states are $V_1=1.95\pm0.04$ and $V_2=3.72\pm0.07$ for all realizations. Theoretical predictions are represented by solid (harmonic mean) and dashed (arithmetic mean) curves.}
\label{fig5_expData}
\end{center}
\end{figure}

We now test the universal averaging protocols experimentally using input states with partially correlated noise. The two AMs were used to impose correlated noise onto the two states by employing a joint electronic noise generator. Due to the correlations, the noise interfere at the balanced beam splitter either constructively to produce a highly noisy output state or destructively to produce an output state with reduced noise. The output with reduced noise then directly serves as the output of the arithmetic mean protocol. To execute the harmonic mean, the output with increased noise is measured, post selected with different threshold values and finally used to herald the quadrature data measured with the verifying detector (HD2). The experimental results of the arithmetic and harmonic means for three different correlation coefficients are shown in Fig.~\ref{fig5_expData}a. For all implementations the correlations were biased with the individual variances being $V_1=1.95\pm0.04$ and $V_2=3.72\pm0.07$. We clearly see that as the correlation becomes stronger, both the arithmetic and harmonic mean variances are reduced. This trend is further illustrated in Fig.~\ref{fig5_expData}b where the success probability for the harmonic mean is set to 0.10. In all figures we insert the theoretical predictions (curves) based on the experimental parameters. This clearly demonstrates that the quantum averaging process is improved if the noises are correlated.

In summary, we have extended the notion of arithmetic mean and, in particular, harmonic mean to the field of quantum optics. Several schemes of implementing the arithmetic and harmonic means of quadrature variances of a quantum state of light have been devised. Experimentally, we have demonstrated a probabilistic scheme for the implementation of the harmonic mean based on a measurement-induced operation, and the results have been compared to the results of a trivial arithmetic mean protocol. We found that the harmonic mean protocol is the best transformation for stabilizing squeezed state resources.

It is interesting to note that the stabilization of squeezed states of light using the harmonic mean law can be readily extended to other media, such as squeezing of the collective spin of an atomic ensemble~\cite{kuzmich}, squeezing in Bose-Einstein condensates~\cite{esteve} and squeezing in plasmonic systems~\cite{Huck2009}. As the squeezed state has been shown to be the basic, irreducible off-line resource for universal state preparation~\cite{fiurasek2005,Dakna1999} and universal quantum computation with continuous variables~\cite{Menicucci2006,Filip2005}, we foresee that the harmonic mean protocol might play a central role in future quantum informational and metrological technologies.

{\bf Acknowledgments} This work was supported by the EU project COMPAS (no.212008), the Danish Agency for Science Technology and Innovation (no. 274-07-0509) and Lundbeck fonden. RF also acknowledges MSM 6198959213 and LC06007 of the Czech Ministry of Education and grant 202/08/0224 of GA CR.

\end{document}